\documentclass[aps,prb,preprint,showpacs,floatfix,preprintnumbers]{revtex4}
\usepackage{graphicx}
\usepackage{amssymb}
\begin{document}

\title{Investigation of element-specific and bulk magnetism, electronic and crystal structures of La$_{0.70}$Ca$_{0.30}$Mn$_{1-x}$Cr$_{x}$O$_{3}$}

\author{H. Terashita$^{1,2}$}
\author{J. C. Cezar$^{3}$}
\author{F. M. Ardito$^{1}$}
\author{L. F. Bufai\c{c}al$^{4}$}
\author{E. Granado$^{1,2}$}

\affiliation{$^{1}$Instituto de F\'{i}sica ``Gleb Wataghin'', Universidade Estadual de Campinas - UNICAMP, 13083-859 Campinas, SP, Brazil\\
$^{2}$Laborat\'{o}rio Nacional de Luz S\'{i}ncrotron, 13083-970 Campinas, SP, Brazil\\
$^{3}$European Synchrotron Radiation Facility (ESRF), F-38043, Grenoble, France\\
$^{4}$Instituto de F\'{i}sica, Universidade Federal de Goi\'{a}s, 74001-970, Goi\^{a}nia, GO, Brazil}

\date{\today}

\begin{abstract}

The magnetic interactions in La$_{0.70}$Ca$_{0.30}$Mn$_{1-x}$Cr$_{x}$O$_{3}$ ($x$=0.15, 0.50 and 0.70) are investigated by x-ray absorption spectroscopy (XAS), x-ray magnetic circular dichroism (XMCD), high-resolution x-ray powder diffraction, and bulk magnetization measurements. XAS in the Mn and Cr $L_{2,3}$ edges support stable single valent Cr$^{3+}$ ions and a varying Mn valence state with $x$, while the O $K$ edge XAS spectrum reveals local maxima in the O $2p$ density of states close to the Fermi level due to mixing with Mn and Cr $3d$ states. A robust antiferromagnetic state is found for $x$=0.70 below $T_N$=258 K. For $x$=0.15, combined XMCD and bulk magnetization measurements indicate a fully polarized ferrimagnetic state for the Mn and Cr spins below $T_c$=224 K. For $x$=0.50, a reduced ferrimagnetic component dominated by Mn spins is present below $T_c$=154 K. No evidence of lattice anomalies due to cooperative charge and orbital orderings is found by x-ray diffraction for all samples. The magnetic properties of this system are rationalized in terms of a competition of ferromagnetic Mn-Mn double exchange and antiferromagnetic Cr-Cr and Cr-Mn superexchange interactions.

\end{abstract}


\maketitle

\section{Introduction}

Transition metal oxides exhibit a wide variety of physical phenomena, such as metal-insulator transitions, high-$T_c$ superconductivity, charge and orbital orderings, high-spin/low-spin states, phase separation, and colossal magnetoresistance (CMR).\cite{Review 1}
In particular, the magnetic and transport properties of mixed-valence manganites La$_{1-x}$Ca$_{x}$MnO$_{3}$ are determined by a competition of ferromagnetic double exchange (DE) interaction involving a hopping of $e_{g}$ electrons between Mn ions\cite{DE 1} and tendencies towards insulating ground states with charge and/or orbital ordering of this electron, in which strongly directional superexchange (SE) interactions determine the magnetic ground state. This fine balance can be disturbed by substitutions in the Mn site, such as demonstrated by the CMR effect induced by Cr-, Co-, and Ni-substitution in Pr$_{0.5}$Ca$_{0.5}$MnO$_{3}$ (ref. \cite{Maignan CO}). Despite the significant amount of work performed in manganites with transition-metal substitution, the origin of observed properties for these compounds are not fully understood in many cases due to the additional complexity brought by atomic disorder in the transition metal site.

Cr$^{3+}$ has an isoelectronic structure with respect to Mn$^{4+}$ ($3d^3$, $s$=3/2) and similar ionic radius of Mn$^{3+}$ (Cr$^{3+}$: 0.615 \AA, Mn$^{3+}$: 0.645 \AA).\cite{Shannon 76}
Therefore, substitution of Cr in the Mn site is ideal to study additional DE and/or SE interactions in manganese oxides without changing structural properties dramatically.
The influence of Cr-substitution on the magnetic and transport properties of manganese oxides has been studied in a wide range of Cr-substitution levels. Many studies focused on LaMnO$_{3}$, which is $A$-type antiferromagnetic (AFM) at low temperatures, resulting from orbital ordering and correspondingly anisotropic Mn$^{3+}$/Mn$^{3+}$ SE interaction.\cite{Review 1, DE 1, Goodenough 63}
A small ferromagnetic (FM) component is then induced by substitution of Cr$^{3+}$ and the role of Cr$^{3+}$ ions in weak ferromagnetism has been investigated.\cite{Gundakaram 96, Zhang 00, Sun 01, Deisenhofer 02, Morales 05}
Several groups\cite{Zhang 00, Sun 01, Morales 05} suggested the possibility of DE interaction between Mn and Cr, while Gundakaram $et$ $al.$ reported that the origin of weak ferromagnetism in Cr-doped LaMnO$_{3}$ is a ferromagnetic SE interaction.\cite{Gundakaram 96}
Similarly, a small FM moment has been observed in La$_{1-x}$Ca$_{x}$CrO$_{3}$ and mixed-valence state of Cr was reported.\cite{Sakai 96, JJN 04}
It was also suggested that a possible presence of Cr$^{4+}$ ions induces canting of the Cr AFM spin alignment in LaCrO$_{3}$, leading to a small net magnetic moment.\cite{JJN 04}

Cr-substitution in the well-known CMR compound La$_{0.70}$Ca$_{0.30}$MnO$_{3}$ may be an alternative strategy to investigate magnetic interactions between Mn and Cr ions in a perovskite crystal structure. In such cases, oxygen stoichiometry is less critical for magnetic properties than in Cr-doped LaMnO$_{3}$ samples,\cite{Roosnalen 94} allowing for a more reliable comparison of the results obtained by the different groups. Additional interactions between Mn spins must be taken into account to understand the phase diagram of this system. Based on magnetization and transport measurements on polycrystalline La$_{0.70}$Ca$_{0.30}$Mn$_{1-x}$Cr$_x$O$_{3}$ samples, Cabeza {\it et al.} inferred an AFM coupling between Mn and Cr spins and suggested that Mn$^{3+}$ $e_g$ orbitals point away from the Cr$^{3+}$ and torwards Mn$^{4+}$ neighboring ions, optimizing DE interactions between Mn ions only.\cite{Cabeza 99} On the other hand, Sun {\it et al.} suggested\cite{Sun 00} DE interactions between Mn and Cr, while Ganguly {\it et al.} proposed FM Mn$^{3+}$/Cr$^{3+}$ SE interactions.\cite{Ganguly 00}
A recent neutron powder diffraction study in La$_{0.70}$Ca$_{0.30}$Mn$_{1-x}$Cr$_{x}$O$_{3}$ suggested that the Mn and Cr net moments are entirely independent.\cite{Capogna 08} According to this, only Mn spins get ordered and Cr$^{3+}$ spins act as random magnetic impurities for $x$=0.15, while only Cr$^{3+}$ spins get ordered and Mn spins are randomly oriented even at 5 K for $x$=0.50 and 0.70 (ref. \cite{Capogna 08}). This scenario suggests a weak coupling between Mn and Cr spins in a extended range of Mn/Cr relative occupancy. However, it is important to mention that Mn and Cr occupy the same crystallographic site and that neutron diffraction is only weakly element sensitive, due to the similar but not identical magnetic form factors of Cr and Mn spins.\cite{Capogna 08} Therefore, the relative orientation between Mn and Cr spins in La$_{0.70}$Ca$_{0.30}$Mn$_{1-x}$Cr$_x$O$_{3}$ is not unambiguously established. This fact limits our understanding on the magnetic interactions in this system. In the present work, we conducted soft x-ray absorption spectroscopy (XAS) and x-ray magnetic circular dichroism (XMCD) measurements for La$_{0.70}$Ca$_{0.30}$Mn$_{1-x}$Cr$_{x}$O$_{3}$ ($x$=0.15, 0.50, and 0.70), which are element-specific probes, complemented by bulk magnetization and high-resolution x-ray powder diffraction (XPD) experiments. Informations on the electronic configuration and relative spin orientation of Mn and Cr ions are obtained, leading to a comprehensive scenario where the magnetic ground states and the most important exchange interactions are identified.

\section{Experimental Details}

Polycrystalline samples of La$_{0.70}$Ca$_{0.30}$Mn$_{1-x}$Cr$_{x}$O$_{3}$ with $x$=0.15, 0.50, and 0.70 were prepared using standard solid-state reaction.
Stoichiometric quantities of La$_2$O$_3$ (dried at 700 $^\circ$C), CaCO$_3$, Mn$_2$O$_3$, and Cr$_2$O$_3$ were mixed and heated in air at 900 $^\circ$C for 15 h, 1100 $^\circ$C for 15 h, 1200 $^\circ$C for 15 h, and 1300 $^\circ$C for 36 h with intermediate grindings.
The samples were then reground and pressed into pellets and reacted at 1400 $^\circ$C for 24 h followed by slow cooling at a rate of 1.5 $^\circ$C/min.
Magnetization measurements were conducted with a commercial Superconducting Quantum Interference Device magnetometer during warming after zero-field cooling.

High-resolution XPD measurements were conducted on the XPD beamline of the Brazilian Synchrotron Light Laboratory (LNLS),\cite{Ferreira} using a wavelength of 1.240 {\AA} and 2$\theta$ steps of 0.008$^\circ$ and a reflection ($\theta-2\theta$) geometry. Loose powders were uniformly spread over a flat Cu sample holder, which was attached to the cold finger of a He closed cycle cryostat. A Ge(111) crystal analyzer was mounted upstream the scintillation detector, to optimize instrumental resolution of 2$\theta$. Rietveld analysis was carried out with the program RIETAN-FP.\cite{RIETAN}

XAS and XMCD measurements at the Mn and Cr \textit{L}$_{2, 3}$ and O \textit{K} edges were carried out on the beamline ID08 at the European Synchrotron Radiation Facility (ESRF). Instrumental energy resolution was 0.13 eV at the O \textit{K} and Cr $L$ edges and 0.16 eV at the Mn $L$ edges. In order to obtain uniform and clean surfaces, the pellets were scraped with a diamond file in the preparation chamber under a pressure of $\sim 10^{-8}$ mbar just before the measurements.
The pressure was kept at about $5 \times 10^{-10}$ mbar during measurements.
The spectra were obtained with $\sim 100$\% circularly polarized beam in the total electron yield mode, and normalized by the incident beam intensity. The XMCD data were taken by alternating the ellipticity of the beam between circular left and right, in an applied magnetic field of $1.5$ T along the beam direction by means of a superconducting magnet.

\section{Results and Analysis}

\subsection{Bulk magnetization}

\begin{figure}[h]
\includegraphics[scale=1]{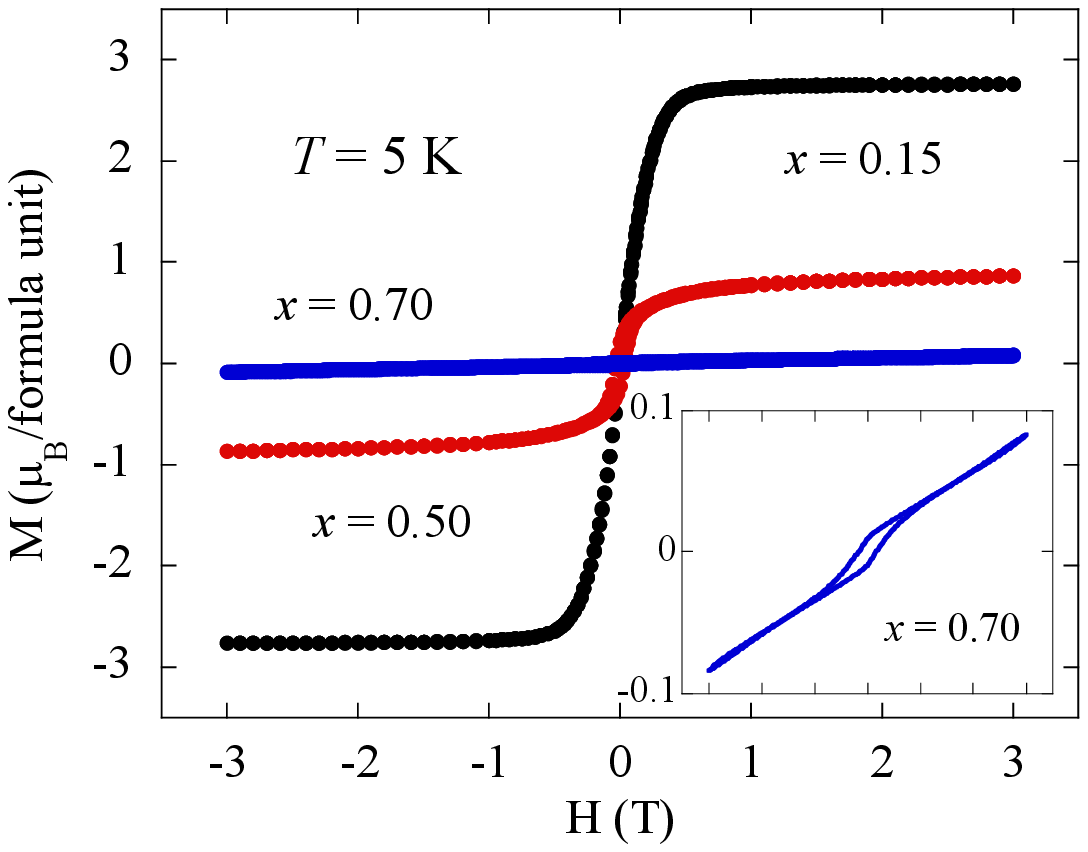}
\caption{\label{magH} (Color online) Magnetization $M$ versus magnetic field $H$ at 5 K for La$_{0.70}$Ca$_{0.30}$Mn$_{1-x}$Cr$_{x}$O$_{3}$ with $x$=0.15, 0.50, and 0.70 after zero field cooling. The inset shows the data for $x$=0.70 in an expanded $M$ scale.}
\end{figure}

Figure \ref{magH} shows the field dependence of $dc$-magnetization $M(H)$ for La$_{0.70}$Ca$_{0.30}$Mn$_{1-x}$Cr$_{x}$O$_{3}$ with $x$=0.15, 0.50, and 0.70 at $T$=5 K. For $x$=0.15 and 0.50, $M(H)$ shows a marked FM component and reaches the values of about 2.8 and 0.8$\mu_B$/formula unit in $H$=3 T, respectively. These values are significantly smaller than the spin-only magnetic moment\cite{DE 1} $M_0$=3.7$\mu_B$/formula unit for the parent compound La$_{0.70}$Ca$_{0.30}$MnO$_{3}$. For $x$=0.70, $M(H)$ shows a dominant linear component and a weak ferromagnetism of 0.029$\mu_B$/formula unit, as seen in the hysteresis loop in the inset of Fig. \ref{magH}. These results are in good agreement with previous reports.\cite{Capogna 08, Cabeza 99}

\begin{table}
\caption{\label{tableM} Magnetic moment
$M_s$ at 5 K, magnetic transition temperatures $T_c$ and $T_N$, Curie-Weiss temperature $\theta_{CW}$, and paramagnetic effective moment $p_{eff}$ for La$_{0.70}$Ca$_{0.30}$Mn$_{1-x}$Cr$_{x}$O$_{3}$ (0$\leq $$x$$\leq $1.0).}

\begin{ruledtabular}
\begin{tabular}{cccccccccc}
$x$ & $M_{s}$ ($\mu_B$/f.u.)\footnotemark[1] & $T_{c, N}$\footnotemark[2] (K) & $\theta_{CW}$ (K) \footnotemark[3] & $p_{eff}$ \footnotemark[3]
\\
\hline
0\footnotemark[4] & 3.38 & 268 \\
0.15 & 2.72 & 224 & 222 & 5.38 \\
0.50 & 0.70 & 154 & 126 & 3.62 \\
0.70 & 0.029 & 258 & -270 & 3.35 \\
1.00\footnotemark[4] & - & 0.002 & 209 \\
\end{tabular}
\end{ruledtabular}

\footnotetext[1] {Straight lines are drawn through the linear portion of the low-field and the high-field magnetization at 5 K. The intercept of these lines are defined as $M_{s}$ for comparison.}

\footnotetext[2] {For comparison, $T_c$ and $T_N$ are defined as intersection point of two linear lines drawn through the $M$ versus $T$ curve above and below the transition in this study. One might determine these temperatures by other methods and/or measurements with similar results.}

\footnotetext[3] {$\theta_{CW}$ and $p_{eff}$ were obtained from fits of inverse magnetic susceptibility data for 300 K$<T<$350 K and $H$=0.4 T.}

\footnotetext[4] {From previous works.\cite{JJN 04, HT 08, HT 01}}

\end{table}

\begin{figure}[h]
\includegraphics[scale=1]{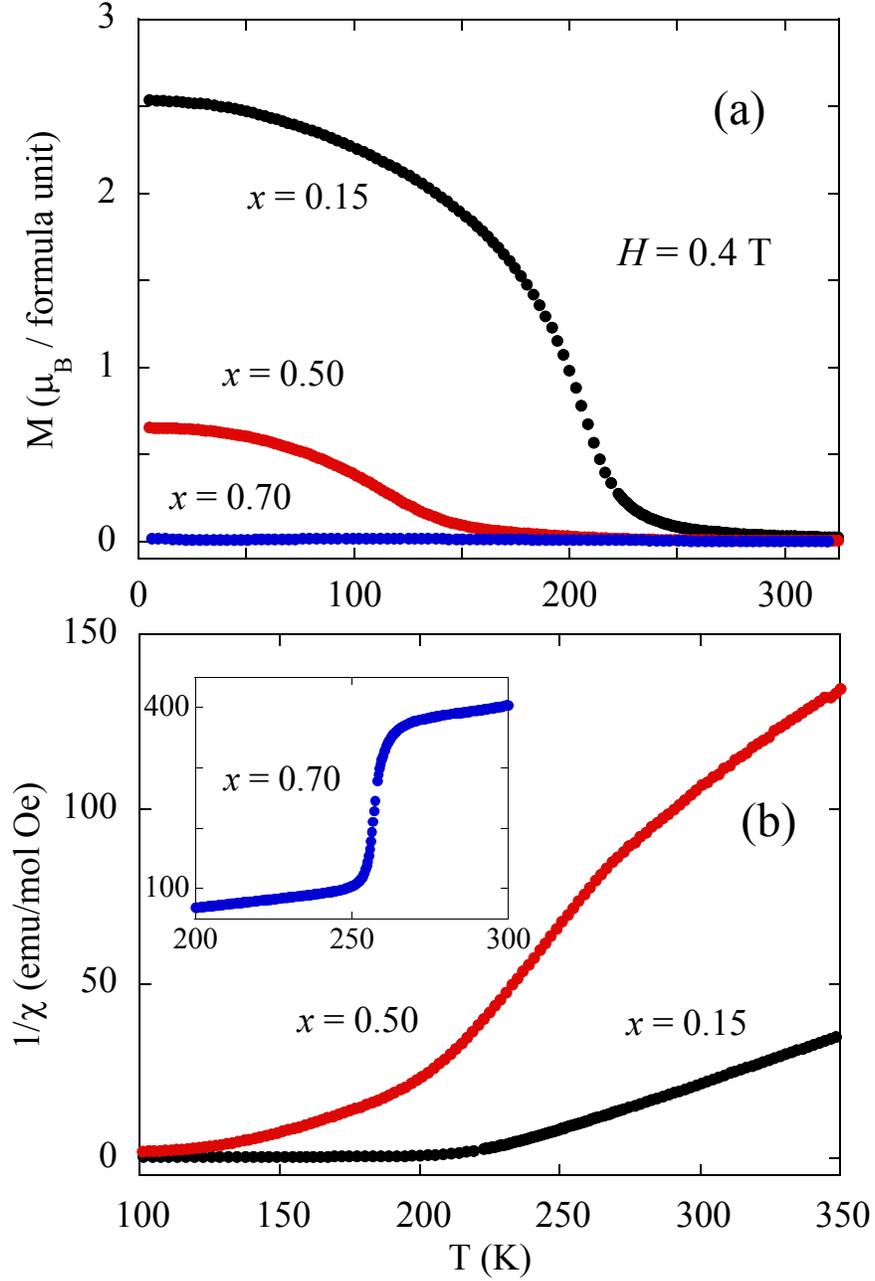}
\caption{\label{magT} (Color online) (a) Temperature dependence of $M$ for $x$=0.15, 0.50, and 0.70 in a magnetic field of $H$=0.4 T after zero field cooling. (b) Inverse magnetic susceptibility 1/$\chi$ versus $T$ around magnetic transition temperature for $x$=0.15 and 0.50. The inset shows 1/$\chi$ versus $T$ for $x$=0.70.}
\end{figure}

The temperature dependence of the magnetization $M$(T) in $H$=0.4 T is shown in Fig. \ref{magT}(a) for all studied samples.
The inverse magnetic susceptibility 1/$\chi$(T) is shown in Fig. \ref{magT}(b). Broad FM transitions, characteristic of CMR-related manganites,\cite{deTeresa 97} are observed at around $T_c$=224 and 154 K for $x$=0.15 and 0.50, respectively. For $x$=0.70, a relatively sharp transition is observed at 258 K. It is inferred from Figs. \ref{magH} and \ref{magT}(b) that the ground state of the $x$=0.70 sample is AFM with a weak FM component, in agreement with neutron powder diffraction.\cite{Capogna 08}

Table \ref{tableM} shows\cite{PM} the transition temperatures $T_c$ or $T_N$, as well as the Curie-Weiss temperatures $\theta_{CW}$ and paramagnetic effective moments $p_{eff}$ obtained from fits of  1/$\chi(T)$ to the Curie-Weiss law for 300 K$<T<$350 K. The large decrease of $\theta_{CW}$ from $+222$ K for $x$=0.15 to $-270$ K for $x$=0.70, clearly indicates that the increasing Cr substitution up to $x$=0.70 weakens the FM interactions in La$_{0.70}$Ca$_{0.30}$Mn$_{1-x}$Cr$_x$O$_3$, favoring the AFM coupling.

\subsection{X-ray powder diffraction}

\begin{table}[t]
\caption{\label{parameters} Structural parameters for La$_{0.70}$Ca$_{0.30}$Mn$_{1-x}$Cr$_{x}$O$_{3}$ at room temperature obtained from Rietveld refinements together with R-factors and goodness of fit for refinements.
Statistical uncertainty is shown in parentheses.}

\begin{ruledtabular}
\begin{tabular}{cccc}
\text{$x$} & \text{0.15} & \text{0.50} & \text{0.70}\\
\text{Space group} & \textit{Pnma} & \textit{Pnma} & \textit{Pnma}\\
\cline{1-4}
\text{$a$ (\AA)} & 5.46028(4) & 5.43994(5) & 5.43248(3)\\
\text{$b$ (\AA)} & 7.71752(5) & 7.69426(6) & 7.68540(6)\\
\text{$c$ (\AA)} & 5.47757(3) & 5.46163(4) & 5.45469(4)\\
\text{$V$ (\AA $^{3}$)} & 230.824(3) & 228.604(3) & 227.737(3)\\
\text{La/Ca} $(x,1/4,z)$\\
\text{$x$} & 0.0189(2) & 0.0168(2) & 0.0174(1)\\
\text{$z$} & 0.9967(4) & 0.9969(5) & 0.9970(3)\\
\text{$U_{iso}$ $\times$ 100 (\AA$^{2}$)}   & 0.83(2) & 0.99(2) & 0.59(2)\\
\text{Mn/Cr} $(0,0,1/2)$\\
\text{$U_{iso}$ $\times$ 100 (\AA$^{2}$)} & 0.59(3) & 0.44(3) & 0.36(3)\\
\text{O(1)} $(x,1/4,z)$\\
\text{$x$} & 0.4936(17) & 0.4941(21) & 0.4960(15)\\
\text{$z$} & 0.0612(18) & 0.0556(22) & 0.0646(18)\\
\text{$U_{iso}$ $\times$ 100 (\AA$^{2}$)} & 0.81(27) & 1.01(34) & 0.67(26)\\
\text{O(2)} $(x,y,z)$\\
\text{$x$} & 0.2731(19) &  0.2761(23) & 0.2737(15) \\
\text{$y$} & 0.0366(9)  &  0.0345(12) & 0.0319(10) \\
\text{$z$} & 0.7230(16) &  0.7233(22) & 0.7257(15) \\
\text{$U_{iso}$ $\times$ 100 (\AA$^{2}$)} &  0.66(18) & 1.30(24) & 1.09(19) \\
\text{Mn(Cr)-O(1) (\AA)} & 1.959(2)  & 1.948(2)  & 1.952(2) \\
\text{Mn(Cr)-O(2) (\AA)} & 1.948(10) & 1.952(13) & 1.954(9) \\
\text{Mn(Cr)-O(2) (\AA)} & 1.980(10) & 1.960(13) & 1.946(9) \\
\text{Mn(Cr)-O(1)-Mn(Cr) ($^\circ$)} & 160.2(6) & 162.0(7) & 159.2(6) \\
\text{Mn(Cr)-O(2)-Mn(Cr) ($^\circ$)} & 159.9(4) & 160.3(5) & 161.9(4) \\
\cline{1-4}
\text{$R_p$} (\%) & 12.93 & 15.42 & 15.24\\
\text{$R_{wp}$} (\%) & 18.46 & 22.23 & 24.61\\
\text{$\chi^2$} & 1.27 & 1.24 & 1.35 \\

\end{tabular}
\end{ruledtabular}
\end{table}

\begin{figure}
\includegraphics[scale=0.30]{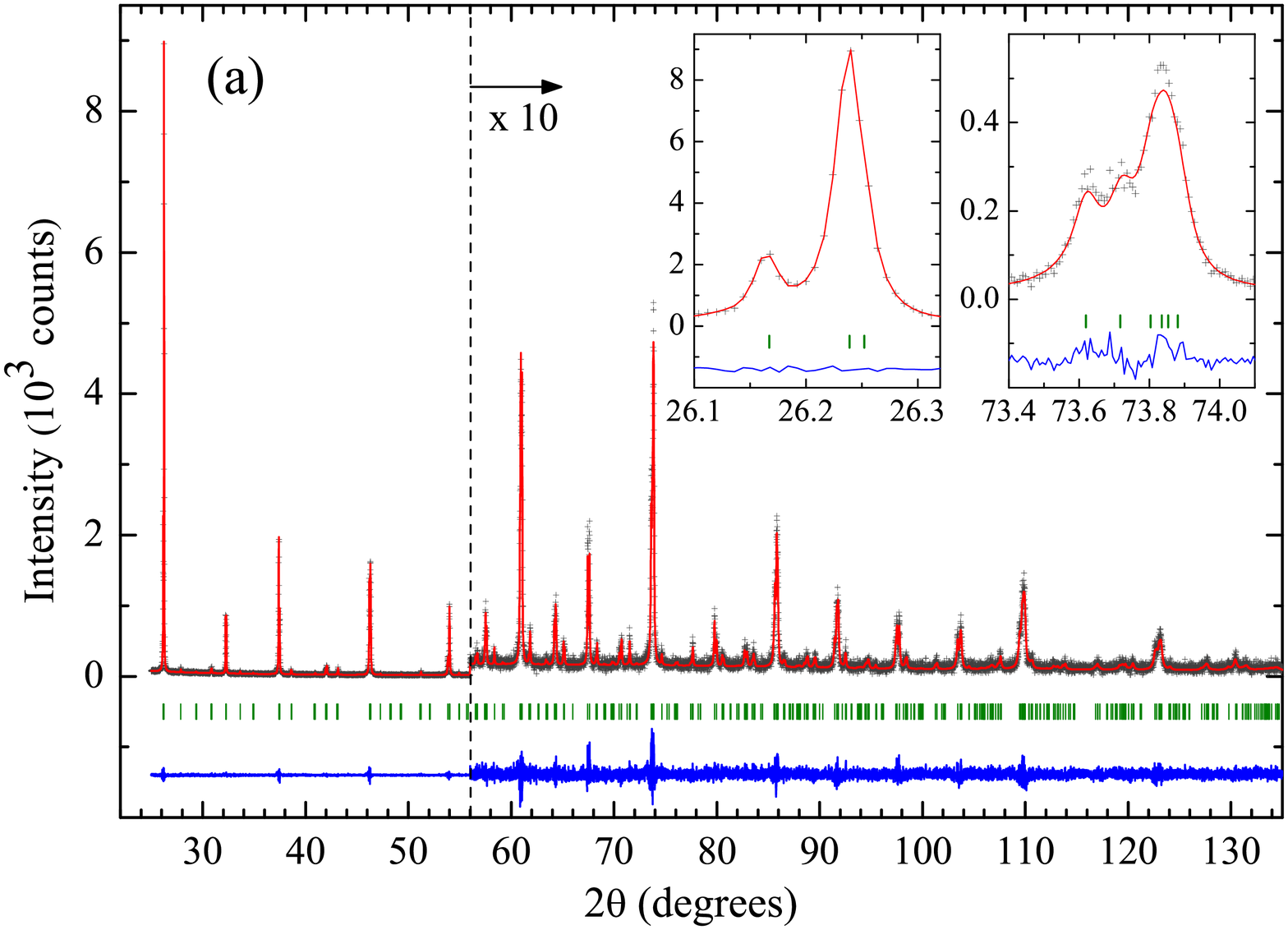}
\includegraphics[scale=0.30]{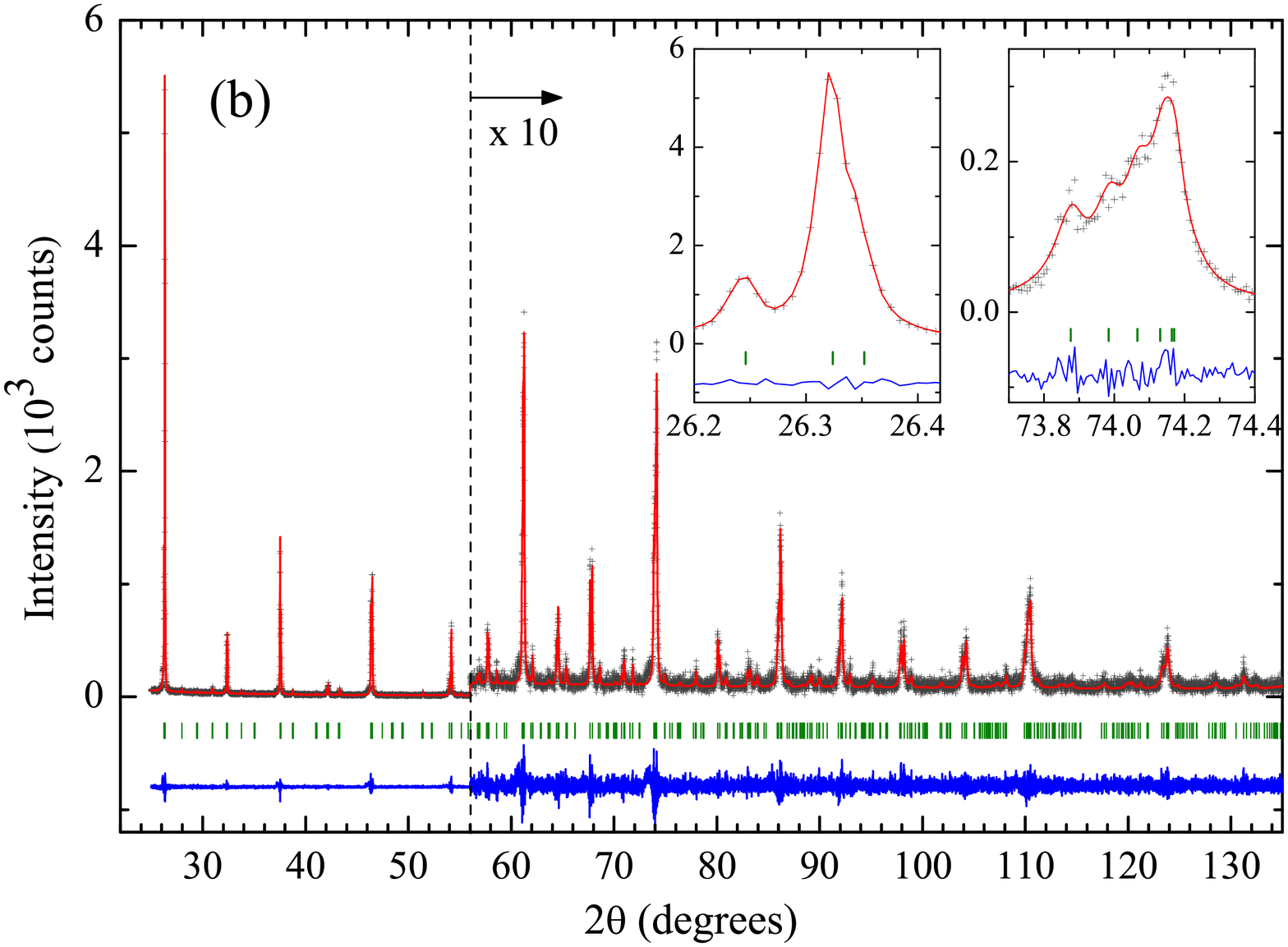}
\includegraphics[scale=0.30]{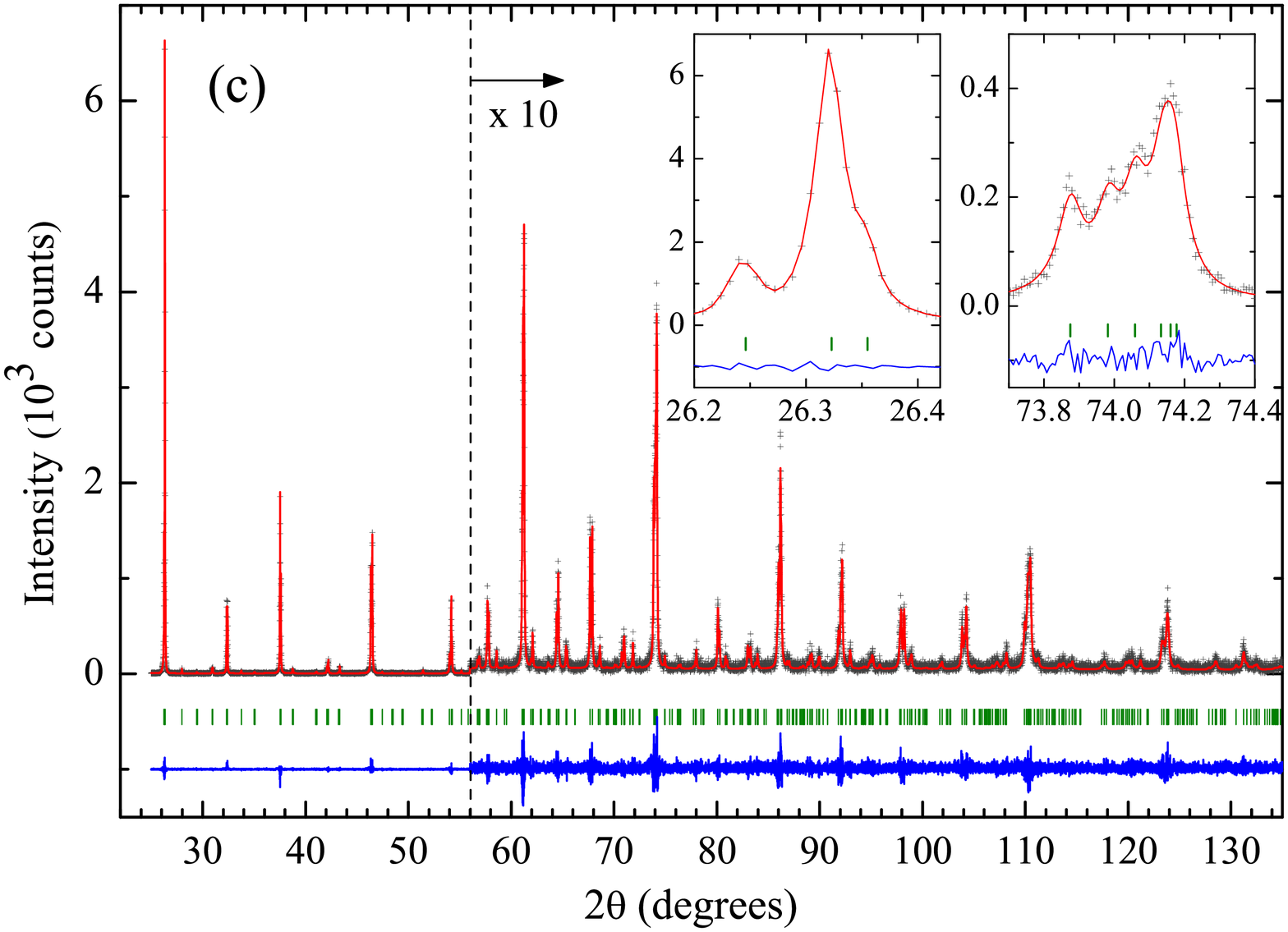}
\caption{\label{pattern} (Color online) Room temperature high resolution x-ray powder diffraction patterns of La$_{0.70}$Ca$_{0.30}$Mn$_{1-x}$Cr$_{x}$O$_{3}$ for (a) $x$=0.15, (b) $x$=0.50, and (c) $x$=0.70 ($\lambda = 1.240$ {\AA}). The cross symbols and solid line represent observed and calculated profiles, respectively. The difference curve is shown at the bottom. Vertical lines indicate expected Bragg peak positions. The insets show selected portions of the profile.}
\end{figure}

\begin{figure}
\includegraphics[scale=1]{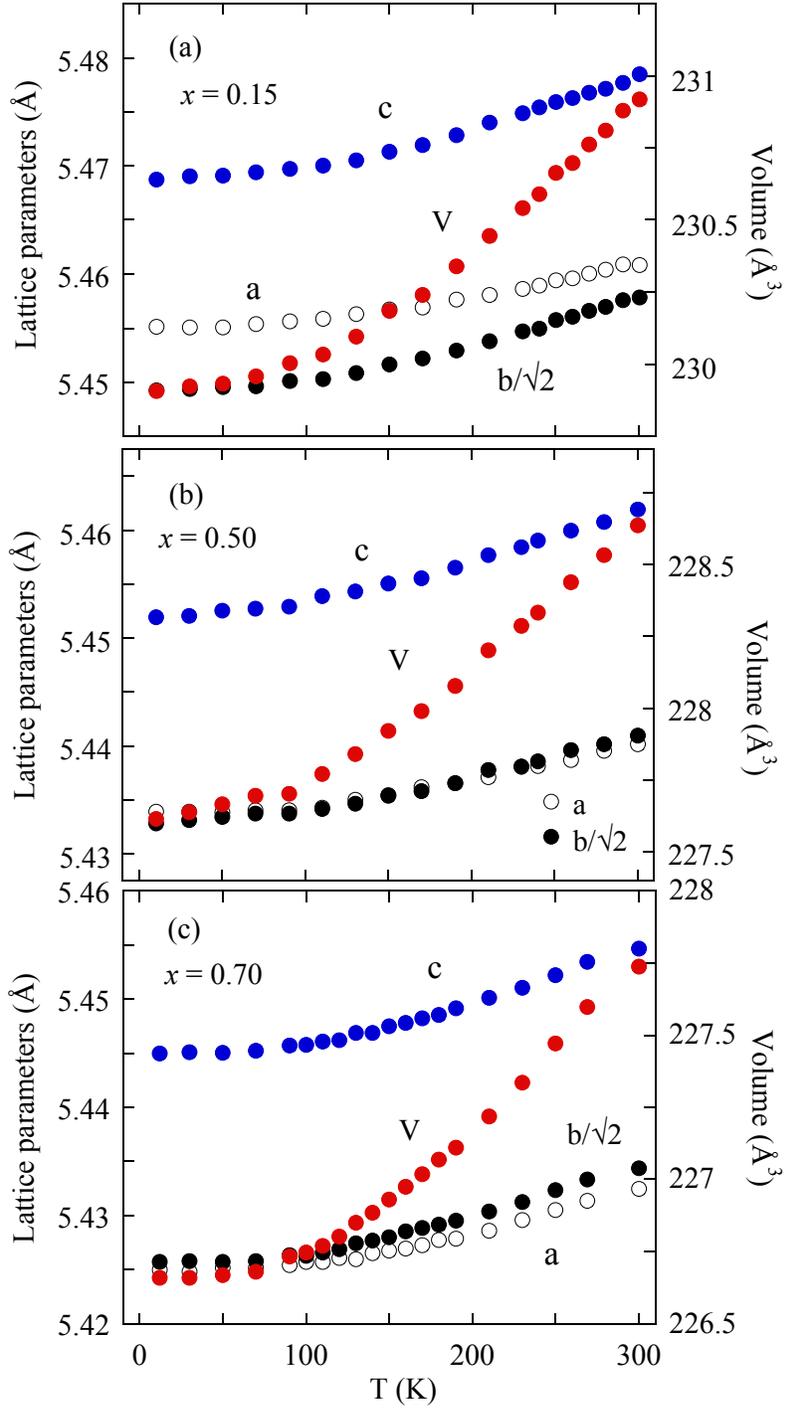}
\caption{\label{lattpar} (Color online) Temperature dependence of the $a$, $b$, and $c$ lattice parameters and unit cell volume $V$ for La$_{0.70}$Ca$_{0.30}$Mn$_{1-x}$Cr$_{x}$O$_{3}$ with (a) $x$=0.15, (b) $x$=0.50, and (c) $x$=0.70.}
\end{figure}

The room temperature x-ray powder diffraction patterns of La$_{0.70}$Ca$_{0.30}$Mn$_{1-x}$Cr$_{x}$O$_{3}$ with $x$=0.15, 0.50, and 0.70 are presented in Figs. \ref{pattern}(a-c). The high angular resolution obtained in our experiment minimizes the superposition of nearby Bragg peaks (see insets of Figs. \ref{pattern}(a-c)). The crystal structures of the studied compounds were refined under a single phase orthorhombic perovskite model (space group $Pnma$). The structural parameters obtained from the refinements at room temperature are given in Table \ref{parameters}. The bond lengths of Mn(Cr)-O in the MnO$_{6}$ octahedra are isotropic and the bond angles of Mn(Cr)-O-Mn(Cr) are nearly the same for all samples. Figures \ref{lattpar}(a-c) show the temperature dependence of the lattice parameters and unit cell volume for the studied compounds. No anomaly in the lattice parameters indicative of cooperative charge and/or orbital ordering of Mn $e_g$ electrons was observed for 10 K$<T<$300 K.

\subsection{X-ray absorption spectroscopy}

\begin{figure}
\includegraphics[scale=1]{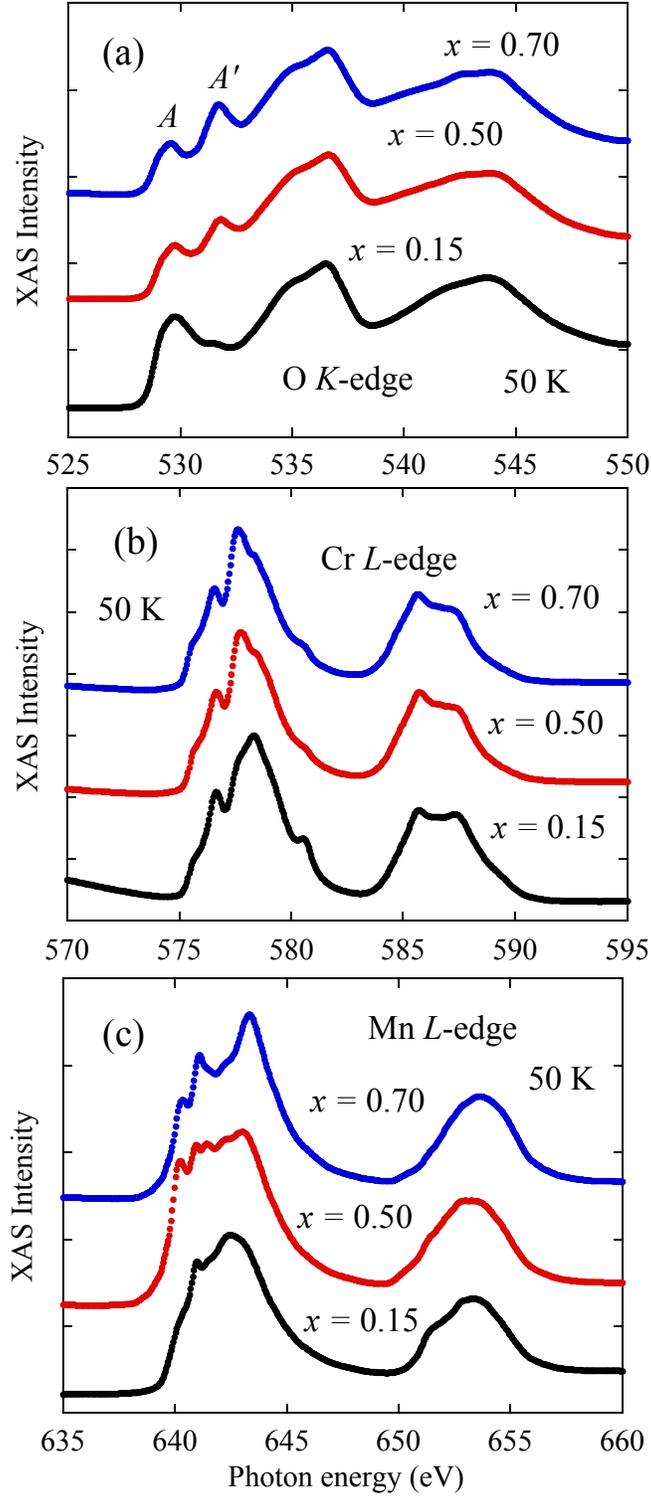}
\caption{\label{XAS} (Color online) Normalized x-ray absorption spectroscopy (XAS) data at the (a) O $K$, (b) Cr $L_{2,3}$ and (c) Mn $L_{2,3}$ edges for La$_{0.70}$Ca$_{0.30}$Mn$_{1-x}$Cr$_{x}$O$_{3}$ with $x$=0.15, 0.50, and 0.70 at 50 K.}
\end{figure}

Figure \ref{XAS}(b) shows the XAS spectra at the Cr $L_{2,3}$ edges for the studied samples at 50 K, probing electronic transitions from the Cr 2$p$ core levels to Cr $3d$ states above the Fermi level.
Similar XAS spectra were observed at the Cr $L_{2,3}$ edges for all studied compounds with four prominent XAS peaks at 576.7, 577.5, 585.6, and 587.5 eV and shoulders at $\sim 576$ and $\sim 580$ eV. These observations agree with published data for Cr$^{3+}$ in CrO$_6$ octahedral environment.\cite{Cr2O3, CuCrO2} Considering the $L_{2,3}$ edge XAS spectra of Cr in other valence states are radically different to the spectra shown in Figure \ref{XAS}(b),\cite{Theil} we conclude that the valence state of Cr is 3+ for all studied samples.

Figure \ref{XAS}(c) shows the XAS spectra at the Mn $L_{2,3}$ edges for the studied samples at 50 K. Contrary to the Cr $L_{2,3}$ edges, large changes in the spectra are noticed with varying Cr/Mn ratio, indicating the electronic modifications with $x$ occur mainly in the Mn ions. The spectrum for $x$=0.70 shows three sharp peaks in the $L_{3}$ edge at 640, 641, and 644 eV and a broad structureless peak in the $L_2$ edge centered at 654 eV. Except for the first small peak at 640 eV, this spectrum is similar to those reported \cite{Sanchez, Abbate 92} for CaMnO$_3$ and La$_{0.1}$Sr$_{0.9}$MnO$_3$, where the formal Mn valence is equal or close to 4+. In fact, it is expected from the charge neutrality of the formula unit that the average Mn state should be 4+ for $x=0.70$, considering the stable Cr$^{3+}$ state demonstrated above. The extra peak at 640 eV is likely related to crystal-field effects caused by the randomly placed Cr$^{3+}$ neighboring ions not present in the above mentioned reference compounds. In the other extreme, the spectrum for $x$=0.15 shows similar line shape of previously reported data for lightly Sr-doped LaMnO$_{3}$.\cite{Abbate 92} It is also important to notice that the highest peak position at the Mn $L_3$ edge shifts from 642.5 eV for $x$=0.15 to 643.3 eV for $x$=0.70. It has been reported for other manganese oxides\cite{Abbate 92, Mitra 03} as well as other transition metal compounds\cite{L shift} that $L$ edge peaks shift to higher energy with increasing the valence state of corresponding ions. 

The XAS spectra at the O $K$ edge probe electronic transitions from the O 1$s$ core level to unoccupied O $p$ states that are hybridized with transitio-metal $3d$ states (see, for instance, ref. \cite{de Groot 89}).
Figure \ref{XAS}(a) shows the XAS spectra at the O $K$ edge for $x$=0.15, 0.50, and 0.70 at 50 K.
The two relatively sharp peaks are observed just below the O $K$ edge at 529.5 eV and 531.5 eV, denoted as $A$ and $A'$. With increasing $x$ from 0.15 to 0.50 and 0.70, a transfer of spectral weight occurs from $A$ to $A'$. For the assignment of these features, we follow the detailed work and comparable results obtained by Toulemonde {\it et al.} in which the O $K$ edge XAS spectra of Sm$_{0.5}$Ca$_{0.5}$Mn$_{1-y}$Cr$_y$O$_3$ ($0.05 \leq y \leq 0.12$) were investigated.\cite{Toulemonde 00} In their work, the $A$ peak was associated with mixing of O $2p$ states with Mn spin up $e_g$ states near the Fermi level, while $A'$ arises from the combined effect of O $2p$ mixing with empty Cr $e_g$ states parallel to the Cr spin, as well as antiparallel Cr $t_{2g}$ states and Mn spin down $t_{2g}$ levels.\cite{Toulemonde 00} The remaining mixing with the Cr antiparallel $e_g$ levels presumably contributes to spectral weight at energies above the $A'$ peak. This assignment by Toulemonde {\it et al.}\cite{Toulemonde 00} seems to be qualitatively consistent with our data, particularly with the clear enhancement of the $A'$ peak with increasing $x$.

The temperature dependencies of the XAS data at all studied edges do not show any relevant modification between 10 K and 300 K. This indicates no electronic phase transition changing the balance between Mn and Cr electronic states all over investigated temperature region.

\subsection{X-ray magnetic circular dichroism}

\begin{figure}
\includegraphics[scale=.7]{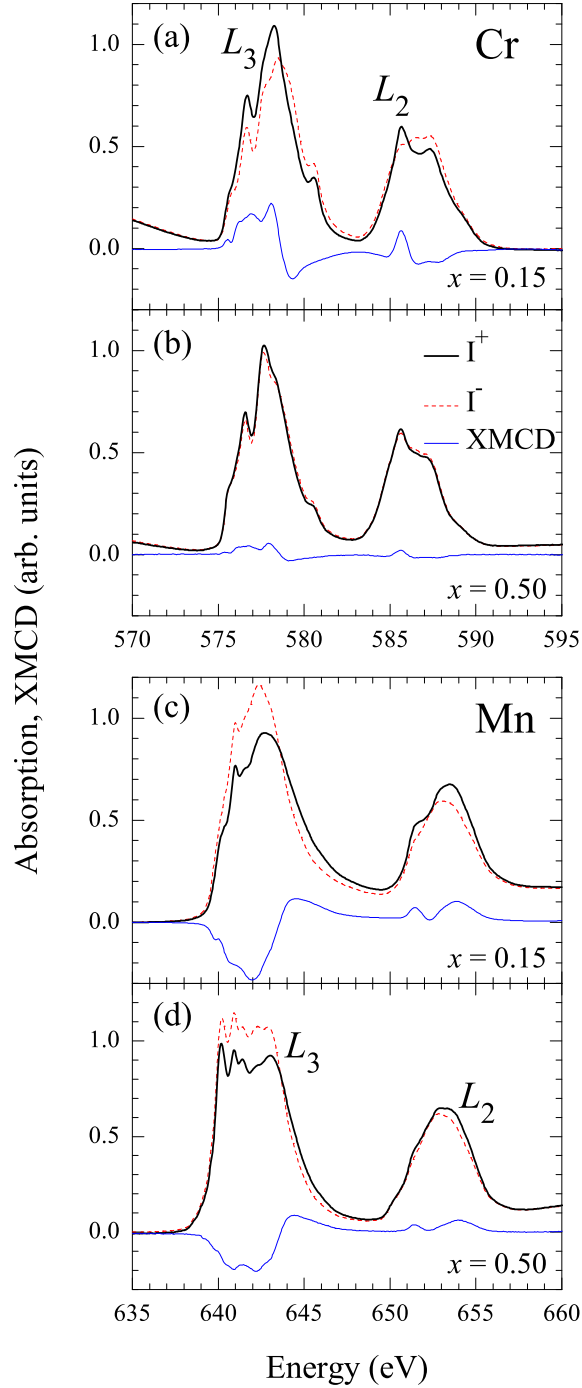}
\caption{\label{XASXMCD} (Color online) X-ray absorption spectra at 10 K with helicity parallel and antiparallel to the applied magnetic field ($I^{+}$ and $I^{-}$, respectively), and corresponding x-ray magnetic circular dichroism (XMCD) signal, $I^{+} - I^{-}$, for La$_{0.70}$Ca$_{0.30}$Mn$_{1-x}$Cr$_{x}$O$_{3}$ with (a) $x$=0.15 at the Cr $L_{2,3}$ edges, (b)  $x$=0.50 at the Cr $L_{2,3}$ edges, (c) $x$=0.15 at the Mn $L_{2,3}$ edges, and (d) $x$=0.50 at the Mn $L_{2,3}$ edges.}
\end{figure}

Figures \ref{XASXMCD}(a,b) show the XAS spectra for $x$=0.15 and 0.50, respectively at 10 K with the beam helicity parallel and antiparallel to the applied magnetic field ($I^{+}$ and $I^{-}$) and the XMCD signal ($I^{+} - I^{-}$) at the Cr $L_{2,3}$ edges. The spectra were normalized with respect to the amplitude of the XAS signal ($I^{+}$ + $I^{-}$)/2. Figures \ref{XASXMCD}(c,d) show the corresponding $I^{+}$, $I^{-}$ and XMCD at the Mn $L_{2,3}$ edges. XMCD signals can be noticed for both samples at the Mn and Cr $L$ edges, indicating that both elements contribute to the magnetic response of these samples at low temperatures. All displayed spectra show a rich structure with a number of characteristic sharp features. On the other hand, the $x$=0.70 sample does not show any relevant XMCD signal (therefore, not shown) due to the absence of a significant FM component (see Fig. \ref{magH}).

\begin{figure}[t]
\includegraphics[scale=1]{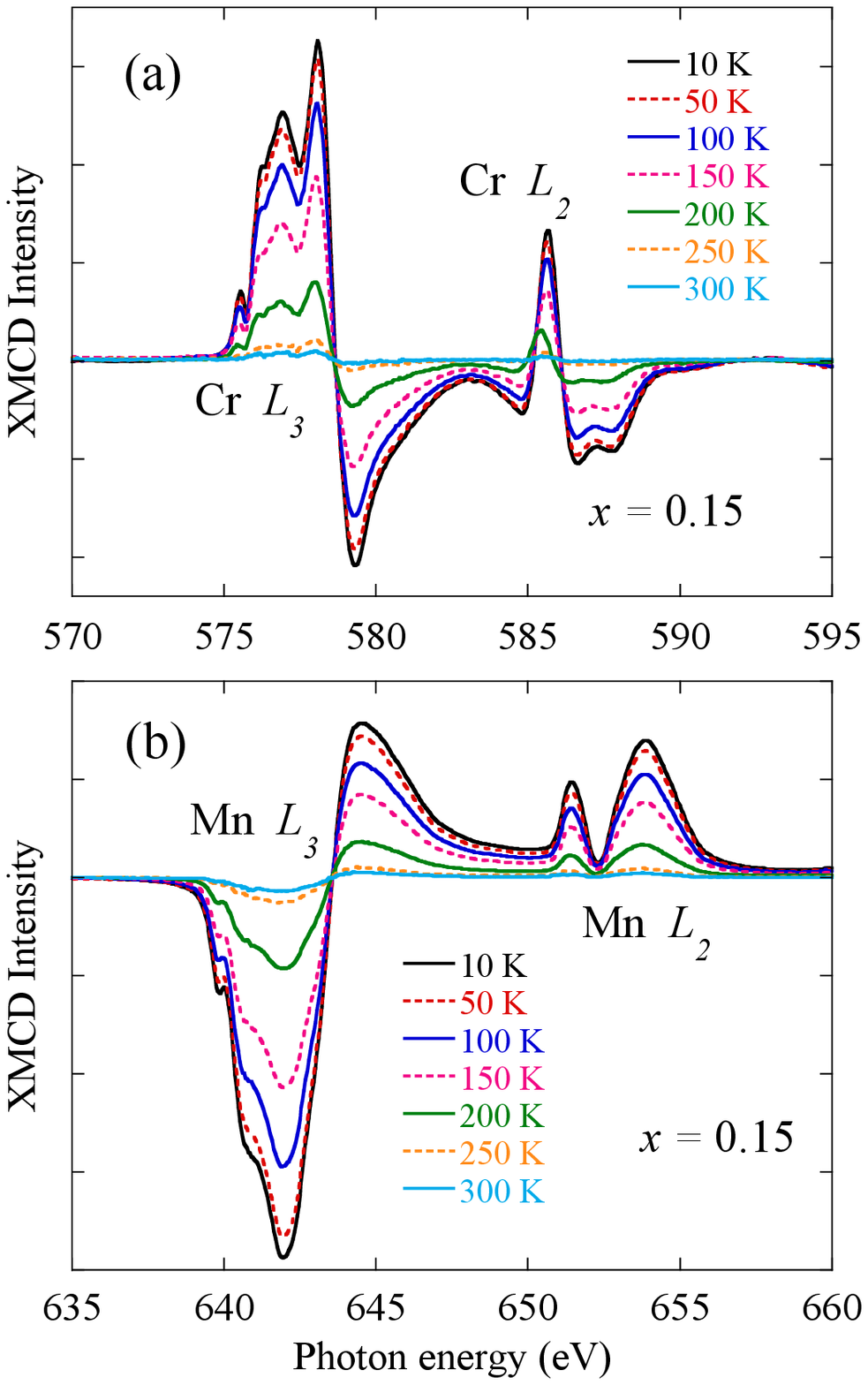}
\caption{\label{XMCD15} (Color online) X-ray magnetic circular dichroism (XMCD) spectra at several temperatures for the (a) Cr \textit{L}$_{2, 3}$ edges and (b) Mn \textit{L}$_{2, 3}$ edges of La$_{0.70}$Ca$_{0.30}$Mn$_{0.85}$Cr$_{0.15}$O$_{3}$.}
\end{figure}

\begin{figure}[t]
\includegraphics[scale=1]{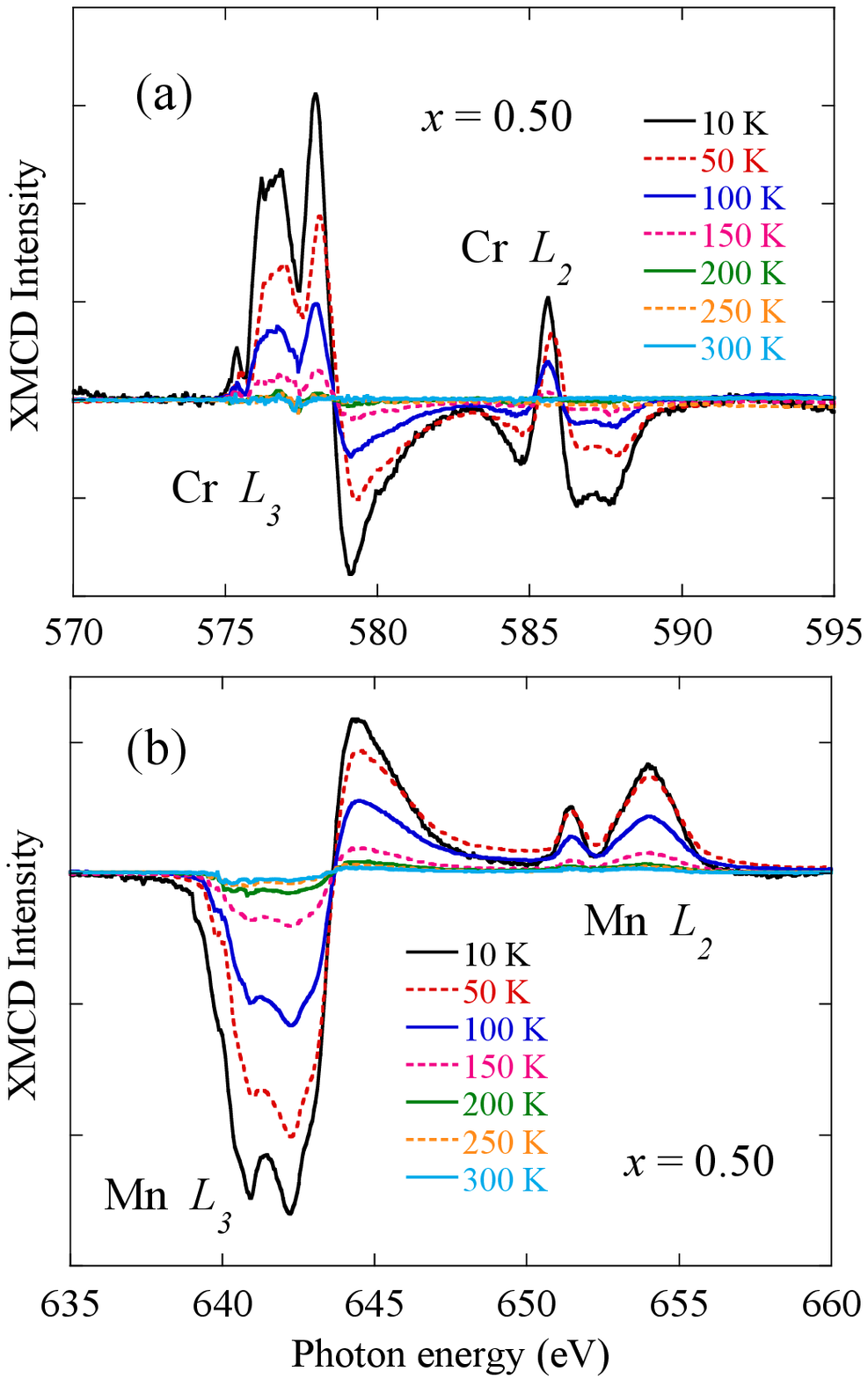}
\caption{\label{XMCD50} (Color online) X-ray magnetic circular dichroism (XMCD) spectra at several temperatures for the (a) Cr \textit{L}$_{2, 3}$ edges and (b) Mn \textit{L}$_{2, 3}$ edges of La$_{0.70}$Ca$_{0.30}$Mn$_{0.50}$Cr$_{0.50}$O$_{3}$.}
\end{figure}

Figures \ref{XMCD15}(a,b) show the detailed XMCD spectra for $x$=0.15 at the Cr and Mn $L_{2,3}$ edges, respectively, at several temperatures, while Figs. \ref{XMCD50}(a,b) show the same for $x$=0.50.
The spectral shape of the XMCD signal is qualitatively similar for both compounds at the  Cr $L_{2,3}$ edges, but distinct at the Mn $L_{3}$ edge, confirming the nearly identical Cr and distinct Mn electronic states for these compounds, as discussed before.

XMCD may provide direct information of $3d$ orbital and spin magnetic moments through sum rules.\cite{Thole 92, Carra 93, Chen 95} However, the application of sum rules to extract the magnetic moments of light transition metals, such as Mn and Cr, has been unsuccessful.\cite{Teramura 96, Scherz 05, Piamonteze 09} Nonetheless, the sign of the Cr and Mn spins with respect to the applied field may be safely inferred from the XMCD integrals. In fact, as seen in Figs. \ref{XMCD15}(b) and \ref{XMCD50}(b), the XMCD area is clearly negative for the Mn $L_3$ edge and positive for the Mn $L_2$ edge for both $x$=0.15 and 0.50, consistent with a positive Mn net FM component with respect to the applied field.\cite{Carra 93, Chen 95} On the other hand, the XMCD area is positive for the Cr $L_3$ edge and negative for the Cr $L_2$ edge, demonstrating that the Cr net FM moment is antiparallel to the applied field and therefore, to the Mn net FM moment for both $x$=0.15 and 0.50. These observations indicate the presence of a ferrimagnetic state of Mn and Cr spins for both compositions.

Due to the similar spectral shapes of the XMCD signal at the Cr $L_{2,3}$ edges for $x$=0.15 and 0.50, the intensity of the Cr XMCD signal for these samples may be directly compared. The amplitude of the XMCD signal for $x$=0.50 at $T$=10 K, defined as the difference between the absolute maximum and minimum of an XMCD spectrum at the $L_{2,3}$ edges, is only $\sim 23$\% of that observed for $x$=0.15 at the same temperature (see Figs. \ref{XASXMCD}(a) and \ref{XASXMCD}(b)), indicating a correspondingly smaller Cr net moment for $x$=0.50 at low temperatures. This comparison is, however, less straightforward for the Mn edges due to different spectral shapes in the $L_3$ edge. Still, the XMCD shape at the Mn $L_2$ edge is comparable for $x$=0.15 and 0.50 and the XMCD amplitude at this edge for  $x$=0.50 is $\sim 63$\% of that observed for $x$=0.15.

\begin{figure}[t]
\includegraphics[scale=0.7]{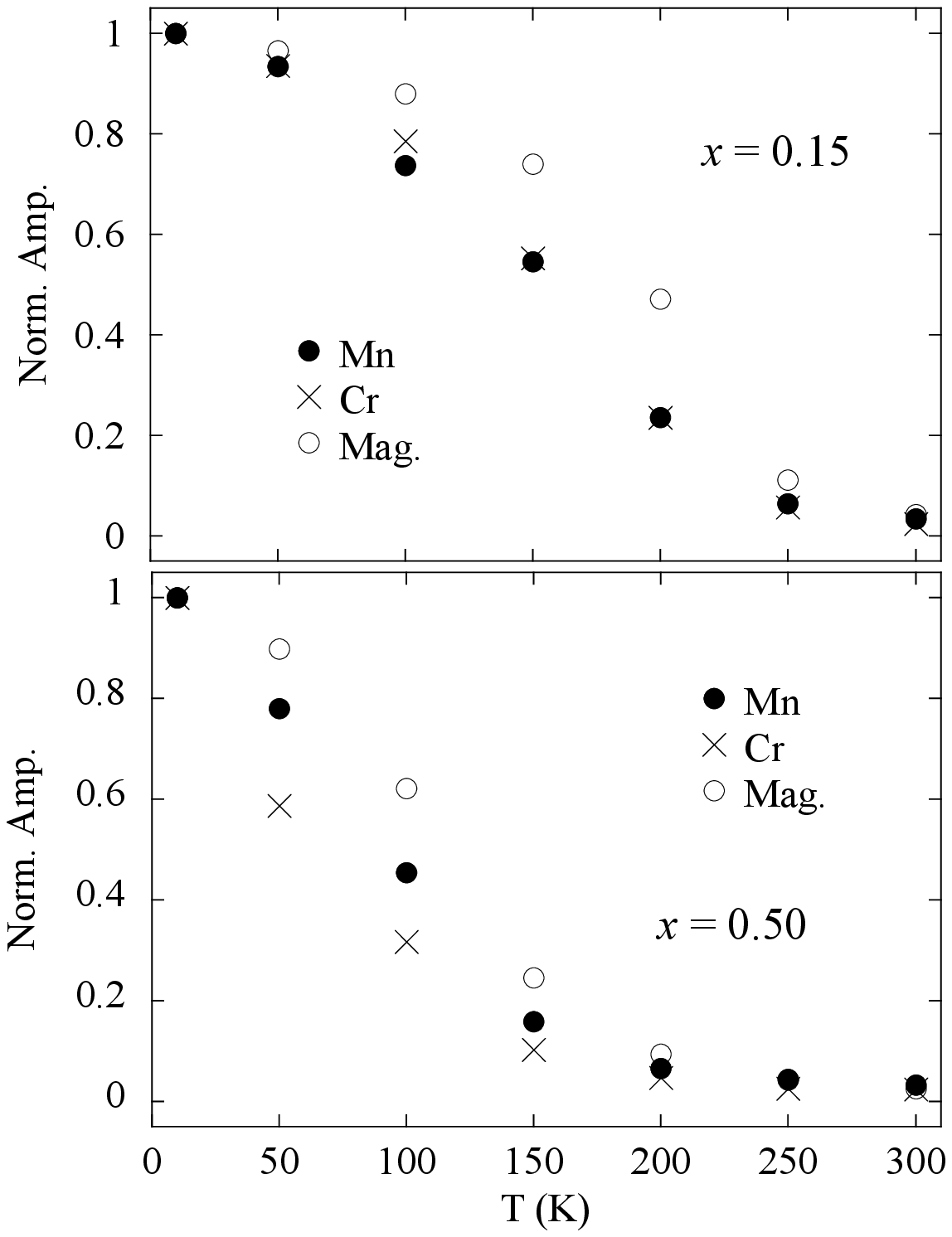}
\caption{\label{max} Temperature dependencies of the amplitudes of the x-ray magnetic circular dichroism (XMCD) signals at the Cr and Mn $L_{2,3}$ edges, normalized at $T$=10 K, for La$_{0.70}$Ca$_{0.30}$Mn$_{1-x}$Cr$_{x}$O$_{3}$ with (a) $x$=0.15 and (b) $x$=0.50. Open circle represent $dc$-magnetization obtained in 1.5 T.}
\end{figure}

In Figures \ref{max}(a,b), the normalized temperature dependencies of the amplitude of the XMCD for the Cr and Mn $L_{2,3}$ edges are presented for $x$=0.15 and 0.50. This amplitude is proportional to the net magnetic moment of the corresponding atom. The normalized Mn and Cr moments fall essentially into the same curve for $x$=0.15, while the Mn and Cr moments for $x$=0.50 show distinct temperature dependencies below $T_c$=154 K. Additional $dc$-magnetization data were taken with the same magneto-thermal sequence of XMCD measurements, $i.e.$ on heating and by alternating magnetic field of 1.5 T and these data are also plotted in Figs. \ref{max}(a,b) for comparison.
The bulk magnetization measurements show different temperature dependencies from those of the XMCD amplitudes as seen in Figures \ref{max}(a,b).
A possible reason for the discrepancy is that XMCD (and XAS) data taken in the total electron yield mode probes the near-surface layers of the sample within a few nanometers, while $dc$-magnetization shows a bulk property.
Nevertheless, the transition temperatures appear to be consistent for both types of measurement in the same magnetic field.

\section{Discussion}

The potential complexity of the magnetic interactions in this system is reduced by the absence of charge and orbital ordering of Mn $e_g$ electrons suggested by our x-ray diffraction data. The absence of charge and orbital ordering in this system is most likely caused by the pinned charge disorder introduced by the random Cr substitution, randomizing Coulomb interactions amongst carriers and locally removing the $e_g$ orbital degeneracy.

The simplest compound in this investigation is La$_{0.70}$Ca$_{0.30}$Mn$_{0.30}$Cr$_{0.70}$O$_{3}$, with only Cr$^{3+}$ and Mn$^{4+}$ ions and therefore, no $3d:e_g$ electrons. Cr$^{3+}$ is electronically equivalent to Mn$^{4+}$ ions with $3d: t_{2g} ^3$ electronic configuration. According to the Goodenough-Kanamori rules,\cite{Goodenough 58} AFM SE interactions between Mn$^{4+}$-Mn$^{4+}$, Cr$^{3+}$-Cr$^{3+}$, and Cr$^{3+}$-Mn$^{4+}$ are expected. In fact, a robust AFM ordering with a sharp transition at $T_N$=258 K is observed in $dc$-magnetization measurement, as shown in Fig. \ref{magT}. This is slightly lower than $T_N$=290 K for LaCrO$_3$ (ref. \cite{Sakai 96}) and much higher than $T_N$=131 K for CaMnO$_3$ (ref. \cite{Neumeier}). A recent neutron powder diffraction study in La$_{0.70}$Ca$_{0.30}$Mn$_{0.30}$Cr$_{0.70}$O$_{3}$ obtained an average magnetic moment of 2.22(2)$\mu_B$ in the Cr/Mn site, significantly smaller than the expected 3$\mu_B$ from simple atomistic considerations.\cite{Capogna 08} It was then suggested that the Mn$^{4+}$ spins act as random magnetic impurities in a matrix of AFM Cr$^{3+}$ spins.\cite{Capogna 08} On the other hand, the relatively high $T_N$ and sharp transition observed here for this compound indicate that both Cr$^{3+}$ and Mn$^{4+}$ ions participate in the magnetic ordering. We suggest that the reduced moments in the transition metal site observed by neutron powder diffraction\cite{Capogna 08} may be due to hybridization between transition-metal $3d$ and oxygen $2p$ states.
It is worth mentioning that even pure LaCrO$_3$ show reduced moments of $\sim 2.5\mu _B$/Cr at low temperatures,\cite{Sakai 96} suggesting this effect is not necessarily related to magnetic disorder.

A weak FM component with a saturation moment of $\sim 0.029$ $\mu_B$ was also observed for $x$=0.70, as shown in Fig. \ref{magH}. It is possible that the origin of small ferromagnetism in $x$=0.70 is a minor impurity of Cr$^{4+}$, as suggested for $x$=0 previously,\cite{JJN 04, Sakai 96} and/or minor impurity of Mn$^{3+}$, which leads to Mn-Mn DE interaction. Antisymmetric Dzyaloshinkii-Moriya interactions, however, cannot be ruled out as one of the possible sources of this weak ferromagnetism.

We proceed with a discussion of the magnetic behavior of the $x$=0.15 sample. As shown in Figs. \ref{XASXMCD}(a) and \ref{XASXMCD}(b), our XMCD data unambiguously demonstrates that the Cr spins are antiparallel with respect to the Mn spins for $x$=0.15, defining a ferrimagnetic state. In addition, as shown in Fig. \ref{max}(a), the element specific Mn and Cr net moments show nearly identical temperature dependencies, indicating that the diluted Cr magnetic moments are strongly coupled to the Mn matrix. According to a simple ferrimagnetic model, the expected bulk magnetization can be calculated as $M_0$=(3.7$-$7$x$)$\mu_{B}$ if the Cr spins are antiparallel to Mn moments,\cite{Eq 1} leading to the predicted moment of 2.65 $\mu_B$/formula unit. This is in good agreement with the observed bulk magnetic moment of 2.72 $\mu_B$/formula unit for $x$=0.15 (see Fig. \ref{magH} and Table \ref{tableM}). On the other hand, this scenario is inconsistent with the random Cr$^{3+}$ spin orientation suggested by Capogna {\it et al.}\cite{Capogna 08} based on an analysis of neutron powder diffraction data.

For the parent compound La$_{0.70}$Ca$_{0.30}$MnO$_{3}$, DE is the dominant magnetic interaction between Mn ions. In La$_{0.70}$Ca$_{0.30}$Mn$_{0.85}$Cr$_{0.15}$O$_{3}$, the relatively small Cr$^{3+}$ concentration allows one to consider these ions as magnetic impurities within a matrix of ferromagnetically aligned Mn spins. The orientation of Cr$^{3+}$ spins with respect to the Mn FM matrix is determined by the SE interactions between Cr$^{3+}$ $t_{2g}$ and Mn $t_{2g}$ and $e_g$ electrons. While the $t_{2g}^{3}-t_{2g}^{3}$ interaction is always AFM,\cite{Goodenough 58} the sign of the Cr $t_{2g}^{3}$ - Mn $e_g$ SE interaction depends on the orientation of the Mn $e_g$ orbitals with respect to the Mn-O-Cr bond direction. The observation reported here of a strong AFM coupling between Mn and Cr spins suggests that the sign of the key $t_{2g}^{3}$ - $e_g$ interaction is also AFM. This is consistent with Goodenough-Kanamori rules\cite{Goodenough 58} if the Mn $e_g$ orbital lobes point along the Mn-O-Mn bonds, and therefore away from the Mn-O-Cr bonds, as originally proposed by Cabeza {\it et al.}\cite{Cabeza 99}

An intermediate scenario with respect to those described above for $x$=0.15 and 0.70 may be drawn for $x$=0.50. In this case of $x$=0.50, the transition metal site is populated with 50\% of Cr and 50\% of Mn. Thus, AFM $t_{2g}^{3}$-$t_{2g}^{3}$ and Cr $t_{2g}^{3}$- Mn $e_g$ SE interactions are also expected to be important for $x$=0.50. These AFM interactions coexist with the FM Mn DE interaction, that are expected to be weaker than for $x$=0.15 due to the smaller concentration of $e_g$ electrons. Our XMCD data suggest the presence of a ferrimagnetic component for both $x$=0.15 and 0.50. Nonetheless, the XMCD amplitudes, presumably proportional to magnetic moment, at the Mn and Cr $L$ edges for $x$=0.50 are significantly smaller than those for $x$=0.15.

Taking the XMCD amplitudes for $x$=0.15 as references for a fully polarized ferrimagnetic state, one may also infer the polarization of Cr and Mn spins in the ferrimagnetic component of $x$=0.50. In fact, the XMCD amplitude at the Mn $L_2$ edge for $x$=0.50 is $\sim 63$\% of that for $x$=0.15. On the other hand, the XMCD amplitude at the Cr $L_{2,3}$ edges for $x$=0.50 is $\sim 23$ \% of that for $x$=0.15. Thus, while a significant fraction of Mn spins are ordered along the same direction for $x=0.50$, only a minority of the Cr magnetic moments are ferrimagnetically ordered. At this point, it is worth questioning where are the remaining magnetic moments for $x=0.50$. To elucidate this point, we mention that an AFM component was previously observed for this composition by neutron diffraction,\cite{Capogna 08} which may coexist with the ferrimagnetic component captured by our XMCD and bulk magnetization data. Due to the relatively large (small) contribution of the Mn (Cr) spins to the ferrimagnetic phase demonstrated by our XMCD data (see above), we conclude that the AFM component for the $x$=0.50 sample is likely composed mostly by Cr spins, in agreement with the conclusions of Ref. \cite{Capogna 08}.

It is note worthy that contrary to $x$=0.15, the normalized Mn and Cr net moments obtained by XMCD for $x$=0.50 do not fall into the same curve as shown in Figs. \ref{max}(a,b), indicating a weaker coupling between the Mn and Cr ferrimagnetic moments. It may be inferred from our XMCD data that the Mn moments for $x$=0.50 are more polarized at temperatures immediately below the magnetic ordering temperature, while the Cr spin polarization becomes relevant only at significantly lower temperatures.

The discussion above for $x$=0.50 is consistent with the scenario drawn above for $x$=0.15, in which the Mn spin polarization is driven by FM Mn DE interactions and the Cr spins are then polarized oppositely to Mn spins by the strong Mn-Cr AFM interactions. The distinction between $x$=0.15 and $x$=0.50 arises from the much higher concentration of Cr ions in the latter. The AFM Cr$^{3+}$-Cr$^{3+}$ interaction then becomes more significant and favors the competing AFM state for the Cr spins against the ferrimagnetic state.

\section{Conclusions}

In summary, we have conducted high-resolution x-ray powder diffraction, magnetization, x-ray absorption, and magnetic circular dichroism measurements to investigate the nature of the exchange interactions in La$_{0.70}$Ca$_{0.30}$Mn$_{1-x}$Cr$_{x}$O$_{3}$ ($x$=0.15, 0.50 and 0.70).
The Cr-Cr and Mn-Cr SE interactions are all AFM, leading to a robust AFM state for $x$=0.70 with $T_N$=258 K. For small Cr concentration ($x$=0.15), Mn-Mn DE leads to a FM polarization of the Mn spins. The Cr spins are then antiferromagnetically coupled to the Mn ones, leading to a fully polarized ferrimagnetic state. Nearly identical temperature dependence of the Mn and Cr net moments for $x$=0.15 reveals a relatively strong AFM coupling between Mn and Cr spins. For $x$=0.50, the significant concentration of Cr spins brings additional AFM Cr-Cr interactions into play, leading to a complex magnetic state with a ferrimagnetic component dominated by Mn spins and an AFM component dominated by Cr spins. The tendency of Mn and Cr spins to order into coexisting ferrimagnetic and antiferromagnetic components for $x \sim 0.50$ explains the peculiar evolution of the bulk saturation moments with $x$, which does not fit into a simple ferrimagnetic model for high Cr concentrations.\cite{Cabeza 99}

\begin{acknowledgments}
This work was supported by the FAPESP and CNPq, Brazil. LNLS is acknowledged for concession of beamtime.
\end{acknowledgments}

\newpage

\newpage

\begin{references}
\bibitem{Eq 1} Spin-only Magnetic moment at $T$=0 K can be calculated as $M_0=2\times\{2\times(0.7-x)+ 3/2\times0.3-3/2x\}\mu_{B}$ for ferrimagnetic specimens. If all Mn and Cr ions are ferromagnetically ordered, $M_0$ for $x$=0.15 can be calculated as $M_0=2\times\{2\times0.65+ 3/2\times0.35\}\times0.85\mu_{B}$+2$\times3/2\times0.15\mu_{B}$, while $M_0$=3.10$\mu_{B}$ is obtained if only Mn ions are ordered.
\end{references}
\end{document}